\documentclass[prd,aps,12pt,preprintnumbers,amssymb,amsmath,amsfonts,nofootinbib]{revtex4-2}

\usepackage{latexsym}
\usepackage{ulem}
\usepackage[utf8]{inputenc}
\usepackage{cancel}

\newcommand{\ba}{\begin{eqnarray}}
\newcommand{\ea}{\end{eqnarray}}

\newcommand{\be}{\begin{equation}}
\newcommand{\ee}{\end{equation}}

\usepackage[x11names,dvipsnames,hyperref]{xcolor}
\usepackage{hyperref}

\usepackage{slashed}
\usepackage{pifont}
\usepackage{dcolumn}
\usepackage{graphics}                                                                         
\usepackage[dvips]{graphicx}
\usepackage{multirow}
\usepackage{makecell}  
\usepackage[labelformat=simple]{subcaption}

\usepackage{tikz}
\usepackage[compat=1.1.0]{tikz-feynman}

\begin{document}
                                                                                
\date{\today}
\title{$X(3872)$, $X(4014)$, and their bottom partners at finite temperature
}

\author{Gl\`oria Monta\~na$^{1,2}$, \`Angels Ramos$^1$, Laura Tolos$^{3,4,5}$ and Juan M. Torres-Rincon$^1$, }

 \affiliation{$^1$Departament de F\'isica Qu\`antica i Astrof\'isica and Institut de Ci\`encies del Cosmos (ICCUB), Facultat de F\'isica,  Universitat de Barcelona, Mart\'i i Franqu\`es 1, 08028 Barcelona, Spain}
 \affiliation{$^2$Theory Center, Thomas Jefferson National Accelerator Facility, 12000 Jefferson Avenue, Newport News, Virginia 23606, USA}
 \affiliation{$^3$Institute of Space Sciences (ICE, CSIC), Campus UAB, Carrer de Can Magrans, 08193, Barcelona, Spain}
 \affiliation{$^4$Institut d'Estudis Espacials de Catalunya (IEEC), 08034 Barcelona, Spain}
 \affiliation{$^5$Frankfurt Institute for Advanced Studies, Ruth-Moufang-Str. 1, 60438 Frankfurt am Main, Germany}
\date{\today}

\begin{abstract}
The properties of the $X(3872)$ and its spin partner, the $X(4014)$, are studied both in vacuum and at finite temperature. Using an effective hadron theory based on the hidden-gauge Lagrangian, the $X(3872)$ is dynamically generated from the $s$-wave rescattering of a pair of pseudoscalar and vector charm mesons. By incorporating the thermal spectral functions of open charm mesons, the calculation is extended to finite temperature. Similarly, the properties of the $X(4014)$ are obtained out of the scattering of charm vector mesons. By applying heavy-quark flavor symmetry, the properties of their bottom counterparts in the axial-vector and tensor channels are also predicted. All the dynamically generated states show a decreasing mass and acquire an increasing decay width with temperature, following the trend observed in their meson constituents. These results are relevant in relativistic heavy-ion collisions at high energies, in analyses of the collective medium formed after hadronization or in femtoscopic studies, and can be tested in lattice-QCD calculations exploring the melting of heavy mesons at finite temperature. 
\end{abstract}

\maketitle


\section{Introduction}
\label{sec:intro}

Over the past decades an incredible amount of charmoniumlike states have been observed experimentally, the so-called $XYZ$. These new discoveries have triggered different theoretical interpretations of their nature, whether they can be understood as tetraquarks, hadroquarkonia states, hadronic molecules, cusps due to kinematic effects, or a mixture of different components (see, for example, the recent reviews~\cite{Guo:2017jvc,Brambilla:2019esw,Liu:2019zoy}).

Among these $XYZ$, the $X(3872)$ [also known as $\chi_{c1}(3872)$] has a prominent role as initiator of the new quarkonium revolution in 2003. First observed in $B^{\pm} \rightarrow K^{\pm} \pi^+ \pi^- J/\psi$ decays by the Belle collaboration~\cite{Choi:2003ue}, its existence  was later confirmed by BABAR~\cite{Aubert:2008gu}, CDF~\cite{Acosta:2003zx,Abulencia:2006ma,Aaltonen:2009vj}, D$\varnothing$~\cite{Abazov:2004kp}, LHCb~\cite{Aaij:2011sn} and CMS~\cite{Chatrchyan:2013cld}. The spin-parity quantum numbers of the $X(3872)$, $J^{PC}=1^{++}$, were extracted at LHCb experiments~\cite{,Aaij:2013zoa,Aaij:2015eva}. This state is quite close to the $D^0 \bar D^{*0}$ threshold, has a width of $\Gamma_X = 1.19 \pm 0.21$ MeV, and is found to decay similarly into $J/\Psi \omega$ and $J/\Psi \rho$ states~\cite{pdg}, signaling to an apparent violation of isospin. The $X(3872)$ can be identified in the  decays of $B$ mesons,
$\Lambda_b$ baryons, as well as in the
radiative decays of charmonia and through lepto- or photoproduction.

In spite of all the available experimental data, the nature of the $X(3872)$ is still evasive. Within constituent quark models, the $X(3872)$ could be understood as a $2\, ^3P_1$ $c \bar c$ charmonium configuration, the $\chi_{c1}(2P)$ state. Nevertheless, the quark model computations give a too large value for the mass of this state  (see, for example, Refs.~\cite{Badalian:1999fe,Barnes:2003vb,Barnes:2005pb})
and other interpretations have appeared. Among them, the $X(3872)$ has been interpreted as a tetraquark state~\cite{Maiani:2004vq,Ebert:2005nc,Matheus:2006xi}  or as a loosely bound hadron molecule~\cite{Tornqvist:2004qy,Wong:2003xk,Thomas:2008ja,Gamermann:2009fv, Gamermann:2009uq,Wang:2013daa}. This latter picture has become very popular due to the closeness of the $D^0 \bar D^{*0}$ threshold, the large decay rate to $D^0 \bar D^{*0}$, as well as the explanation of the apparent isospin symmetry violation in terms of simple phase-space considerations. Other analyses consider the $X(3872)$ as a hadrocharmonium~\cite{Dubynskiy:2008mq}, a mixture between charmonium and exotic molecular states~\cite{Matheus:2009vq, Ortega:2009hj,Cincioglu:2016fkm}, or relate this state with  a  $X$  atom~\cite{Zhang:2020mpi}. We refer the reader to the recent reviews~\cite{Esposito:2014rxa,Lebed:2016hpi,Esposito:2016noz,Guo:2017jvc,Olsen:2017bmm,Kalashnikova:2018vkv,Cerri:2018ypt, Brambilla:2019esw} and references therein.

More recently, the $X(4014)$ state has been observed by the Belle collaboration, appearing as a second structure in the two-photon process $\gamma \gamma \rightarrow \gamma \psi(2S)$~\cite{Belle:2021nuv}. While the first structure seen in  $\gamma \gamma \rightarrow \gamma \psi(2S)$ corresponds to a state with mass $M=3922.4 \pm 6.5 \pm 2$ MeV and width $\Gamma=22 \pm 17 \pm 4$ MeV, and it might be associated to the $X(3915)$, the $\chi_{c2}(3930)$, or a mixture of them, the second state with $M=4014.3 \pm 4 \pm 1.5$ MeV and $\Gamma=4 \pm 11 \pm 6$ MeV could be a new resonance, the $X(4014)$. As discussed in Refs.~\cite{Guo:2013sya,Nieves:2012tt}, the application of heavy-quark spin symmetry
(HQSS) to the charmed meson-antimeson system together with
the identification of the $X(3872)$ and $X(3915)$
states as isoscalar $D \bar{D}^*$ and $D^* \bar{D}^*$ molecular states with $J^{PC}=1^{++}$ and $0^{++}$, respectively, implies the existence of four additional molecular partners. Among those, the $X(4014)$  is identified as a  $D^* \bar{D}^*$ state with $J^{PC}=2^{++}$. Note, however, that the molecular nature of the $X(3915)$ state used as an input in the HQSS analyses is under discussion, as it could be identified as the charmonium $\chi_{c0} (2P)$~\cite{Liu:2009fe,Kher:2018wtv,Duan:2020tsx} or as an $s$-wave $D_s^+ D_s^-$ state~\cite{Li:2015iga}, thus implying that the $X(4014)$ state belonging to the HQSS family is still under debate. 

Whereas the nature of these $X$ states is determined by the charmonium spectrum and the comparison with the branching ratios for two- and three-body decays, the production yield of these exotic hadrons in $pp$ collisions or relativistic heavy-ion collisions (HICs) has opened a new venue of interest. In particular,  there has been a recent controversy on the conclusions about the nature of $X(3872)$ coming from the analysis of the high prompt production cross section of the $X(3872)$ at $pp$ collisions at CDF~\cite{Abulencia:2006ma} and in CMS~\cite{Chatrchyan:2013cld}. The description of the multiplicity dependence of the production rates of promptly produced $X(3872)$ relative to the $\psi(2S)$, observed recently in $pp$ collisions at LHCb~\cite{LHCb:2020sey}, has also lead to debate on the interpretation of the $X(3872)$~\cite{Esposito:2020ywk,Braaten:2020iqw} (see a more detailed discussion and references in~\cite{Albaladejo:2021cxj}). 

One of our motivations to study thermal effects on these hadrons comes from the collective medium formed after hadronization in relativistic HICs. At high collision energies, the eventual hadronic medium is created around temperatures of $T_c=156$ MeV, where interactions among hadrons occur until the so-called kinetic freeze-out at lower temperatures. Among these hadrons, the exotic $X(3872)$ has recently been reconstructed in Pb-Pb collisions by the CMS experiment~\cite{CMS:2021znk}, where a substantial production with respect to $pp$ collisions for the ratio $X(3872)/\psi(2S)$ has been reported. Therefore, this state can suffer from thermal medium modifications during the hadronic expansion until its final decoupling from the medium.

In this context, the $X(3872)$ has already been considered within the statistical hadronization model~\cite{Andronic:2019wva}, incorporated---as the model itself stipulates~\cite{Andronic:2017pug}---with its vacuum properties, i.e., a mass of 3872 MeV and (nearly) zero width. This model shows a prediction for the distribution of transverse momentum, $p_T$, at LHC energies, which is still soon to be tested, and has to wait for future measurements.


The behavior of the charmoniumlike states for extreme conditions, such as those found in HICs at RHIC and LHC energies, is another way to study the nature of these states. As an example, using the coalescence model, the ExHIC collaboration~\cite{Cho:2010db,Cho:2011ew,Cho:2017dcy} has shown that considering the $X(3872)$ as a molecule implies a production yield much larger than for the tetraquark configuration, in particular if one also takes into account the evolution in the hadronic phase~\cite{Cho:2013rpa,Abreu:2016qci}, given that the production and absorption cross sections in HICs are expected to be larger for a molecular state. Indeed, the authors of Ref.~\cite{Abreu:2016qci} corroborated that fact by studying the time evolution of the $X(3872)$ abundance in the hot hadron gas based on all hadronic production mechanisms of the $X(3872)$ assumed in Refs.~\cite{Torres:2014fxa,Abreu:2016qci}. Other approaches have analyzed the nature of the $X(3872)$ in HICs with instantaneous coalescence models~\cite{Fontoura:2019opw,Zhang:2020dwn}, within a statistical hadronization scheme~\cite{Andronic:2019wva}, or using a thermal-rate equation approach~\cite{Wu:2020zbx}.


However, up to now, the analyses of the production of charmoniumlike states, such as the $X(3872)$, in HICs do not take into account their possible in-medium modification in the hot hadronic phase. In the case of the $X(3872)$, there are two recent studies on its in-medium properties. In Ref.~\cite{Cleven:2019cre} the properties of the $X(3872)$ in a finite-temperature pion bath have been studied assuming this resonance to be a molecular state generated by the interaction of $D \bar D^* -\textrm{c.c.}$  pairs and associated coupled channels. Within this approach, the $X(3872)$ develops a substantial width within a hot pionic bath at temperatures close to the critical temperature,  whereas its nominal mass moves above the $D \bar D^*$ threshold. In Ref.~\cite{Albaladejo:2021cxj} the behavior of the $X(3872)$ in dense nuclear matter was studied assuming that  the $X(3872)$ is a purely molecular $(D \bar D^*-\textrm{c.c.})$ state or taking into account mixed-molecular scenarios. Important nuclear corrections for the $D\bar D^*$ amplitude and the pole position of the resonance were found, showing a strong dependence of these results on the $D \bar D^*$ molecular component in the $X(3872)$ wave function, which can be tested in the future experiments PANDA and CBM at FAIR. 

In the present paper we want to pursue the analysis of the nature of the charmoniumlike states by studying the modification due to finite-temperature corrections of the properties of the $X(3872)$ and its possible spin-2 partner, the newly observed $X(4014)$. 
We assume that they are molecular-type bound states of two charmed mesons, generated from the $D^{(*)}\bar{D}^*$ and $D_s^{(*)}\bar{D}_s^*$ coupled-channel interactions, belonging to the same HQSS multiplet. Moreover, we take advantage of the heavy-quark flavor symmetry (HQFS) and analyze the bottom sector. To this aim, we first give predictions for the vacuum masses of $X_b$ bottomoniumlike partners generated from the $B^{(*)}\bar{B}^*$ and $B_s^{(*)}\bar{B}_s^*$ interactions, following the philosophy of Refs.~\cite{Nieves:2012tt, Guo:2013sya}, and then study their modification at finite temperature. 
The heavy-quark flavor partners of the $X(3872)$ have been searched for in the $X_b\rightarrow \pi^+\pi\Upsilon(1S)$ channel from $pp$ collisions by CMS~\cite{CMS:2013ygz} and ATLAS~\cite{ATLAS:2014mka}, and in $e^+e^-\rightarrow\gamma X_b$, $X_b\rightarrow\omega\Upsilon(1S)$ by Belle~\cite{Belle:2014sys} and very recently by Belle-II~\cite{Belle-II:2022xdi}, with no significant evidence found so far.
The final goal of this work is to gain insight into the nature of the $X(3872)$, the $X(4014)$ and the associated bottom partners by analyzing their in-medium properties in hot matter, as a complementary tool to spectroscopic analyses of these charmonium- and bottomoniumlike states.

The paper is organized as follows. In Sec.~\ref{sec:hiddengauge} we present the local hidden-gauge formalism, showing how the pseudoscalar-vector and vector-vector interactions are built. Then, in Sec.~\ref{sec:unitarization} we discuss the unitarization procedure in coupled channels, whereas in Sec.~\ref{sec:temperature} we explain how the unitarization procedure is modified at finite temperature within the imaginary-time formalism. We finalize this paper by presenting our results in Sec.~\ref{sec:results} and our conclusions in Sec.~\ref{sec:conclusions}.




\section{The local hidden-gauge formalism}
\label{sec:hiddengauge}

The interaction between pseudoscalar (P) and vector mesons (V) can be studied using the Lagrangian of the local hidden-gauge symmetry approach~\cite{Bando:1984ej,Bando:1987br,Nagahiro:2008cv},
\begin{equation}\label{eq:LagHG}
 \mathcal{L}=-\frac{1}{4}\langle V_{\mu\nu}V^{\mu\nu}\rangle+\frac{1}{2} m_V^2\Bigg\langle\Bigg(V_\mu-\frac{i}{g}\Gamma_\mu\Bigg)^2\Bigg\rangle \ ,
\end{equation}
where $\langle\cdots\rangle$ stands for the trace in $SU(4)$ flavor space and $V_{\mu\nu}$ is defined as $V_{\mu\nu}=\partial_\mu V_\nu-\partial_\nu V_\mu-i g[V_\mu,V_\nu]$, in terms of the $SU(4)$ matrix $V_\mu$ containing the vector mesons (to be shown later). 
The hidden-gauge coupling constant $g$ is related to the meson decay constant $f$ and the vector meson mass $m_V$ through the relation $g=\frac{m_V}{2f}$, which fulfills the Kawarabayashi-Suzuki-Fayyazuddin-Riazuddin rule of vector-meson dominance~\cite{Kawarabayashi:1966kd,Riazuddin:1966sw}.

The expansion of the Lagrangian of Eq.~(\ref{eq:LagHG}) leads, on one hand, to a four-vector contact term (VVVV),
\begin{equation}\label{eq:LagVVVV}
 \mathcal{L}_{\textrm{VVVV}}^c=\frac{g^2}{2}\langle V_\mu V_\nu V^\mu V^\nu - V_\nu V_\mu V^\mu V^\nu\rangle \ ,
\end{equation}
and, on the other hand, it gives rise to the vertices involving a vector meson and two pseudoscalar mesons (PPV),
\begin{equation}\label{eq:LagPPV}
 \mathcal{L}_{\textrm{PPV}}=-ig\langle V^\mu[P,\partial_\mu P]\rangle \ ,
\end{equation}
and three vector mesons (VVV),
\begin{equation}\label{eq:LagVVV}
 \mathcal{L}_{\textrm{VVV}}=ig\langle(V^\mu\partial_\nu V_\mu - \partial_\nu V_\mu V^\mu)V^\nu\rangle \ .
\end{equation}
We note that there is no contact term for the scattering between pseudoscalar and vector mesons (i.e., PPVV) in the hidden-gauge formalism.

In $SU(4)$, the $P$ and $V^\mu$ matrices collecting the $16$-plets of pseudoscalar- and vector-meson fields can be written as
\begin{equation}\label{eq:matrixP}
P = 
\begin{pmatrix}
\frac{1}{\sqrt{2}}\pi^0+\frac{1}{\sqrt{6}}\eta+\frac{1}{\sqrt{3}}\eta^\prime & \pi^+ & K^{+} &\ \bar{D}^{0} \\
\pi^- & \!\!\!\!\!\!\! -\frac{1}{\sqrt{2}}\pi^0+\frac{1}{\sqrt{6}}\eta+\frac{1}{\sqrt{3}}\eta^\prime & K^{0} & D^{-} \\
K^{-} & \bar{K}^{0} & \!\!\!\!\!\!\!-\sqrt{\frac{2}{3}}\eta+\frac{1}{\sqrt{3}}\eta^\prime & D_s^{-} \\
D^{0} & D^{+} & D_s^{+} & \eta_c \\
\end{pmatrix} \ ,
\end{equation}
and
\begin{equation}\label{eq:matrixVmu}
 V^\mu =
\begin{pmatrix}
\frac{1}{\sqrt{2}}(\rho^0+\omega) & \rho^+ & K^{\ast +} & \bar{D}^{\ast 0} \\
\rho^- & \frac{1}{\sqrt{2}}(-\rho^0+\omega) & K^{\ast 0} & D^{\ast -} \\
K^{\ast -} & \bar{K}^{\ast 0} & \phi & D_s^{\ast -} \\
D^{\ast 0} & D^{\ast +} & D_s^{\ast +} & J/\psi \\
\end{pmatrix}^\mu \ .
\end{equation}

We focus on the sector with hidden charm $C=0$ and strangeness $S=0$ in isospin basis. We consider the $D\bar{D}^*(3875.80)$  and $D_s\bar{D}_s^*(4080.54)$ coupled channels with quantum numbers $I^G(J^{PC})=0^+(1^{++})$, and the 
$D^*\bar{D}^*(4017.11)$ and $D_s^*\bar{D}_s^*(4224.40)$ channels in the case of $I^G(J^{PC})=0^+(2^{++})$, where the numbers in parentheses denote their corresponding energy thresholds in MeV. 
The $K\bar{K}^*(1389.29)$ and $K^*\bar{K}^*(1789.30)$ channels can be safely ignored in this study, given that their threshold masses lie far from the energy region of interest and their contribution to the generation of the $X(3872)$ and the $X(4014)$, respectively, is negligible. We will actually see that the states with hidden strangeness, i.e., $D_s\bar{D}_s^*$ and $D_s^*\bar{D}_s^*$, play also a minor role, and that the $X(3872)$ and the $X(4014)$ are mainly generated from the dynamics of the $D\bar{D}^*$ and $D^*\bar{D}^*$ channels, respectively, due to the proximity of these states to their thresholds. Nevertheless, we consider a coupled-channel basis and allow transitions between different channels for completeness.
We note that we consider positive $G$-parity\footnote{We recall that the $G$-parity operator is defined as $\hat{G}=e^{i\pi T_2}\hat{C}$, with $T_2$ being the second component of the isospin $SU(2)$ operator, and the $C$-parity operator acts on pseudoscalar and vector mesons as $\hat{C}|P\rangle=|\bar{P}\rangle$ and $\hat{C}|V\rangle=-|\bar{V}\rangle$.} (and hence positive $C$-parity) combinations of the PV wave functions in isospin basis, i.e. $|D\bar{D}^*-\textrm{c.c.}\rangle=\frac{1}{2}(|D^0\bar{D}^{*0}\rangle-|\bar{D}^0D^{*0}\rangle +|D^+D^{*-}\rangle - |D^-D^{*+}\rangle)$ and $|D_s\bar{D}_s^*-\textrm{c.c.}\rangle=\frac{1}{\sqrt{2}}(|D_s^+D_s^{*-}\rangle-|D_s^-D_s^{*+}\rangle)$, following the sign convention in~\cite{Gamermann:2009fv}, but employ the simplified notation $D\bar{D}^*$ and $D_s\bar{D}_s^*$ throughout this work.



\begin{figure}[htbp!]
\centering 
\begin{subfigure}[b]{0.32\textwidth}\centering 
\captionsetup{skip=0pt}
 \begin{tikzpicture}[baseline=(i.base)]
    \begin{feynman}
      \vertex (i);
      \vertex [left = 1cm of i] (a) {\(D\)};
      \vertex [right = 1cm of i] (c) {\(D\)};
      \vertex [below = 2cm of i] (j);
      \vertex [below = 2cm of a] (b) {\(\bar{D}^*\)};
      \vertex [below = 2cm of c] (d) {\(\bar{D}^*\)};
      \diagram* {
        (a) -- [charged scalar] (i) -- [charged scalar] (c), 
        (i) -- [edge label={\(\rho,\omega,J/\psi\)}] (j),
        (b) -- [fermion] (j) -- [fermion] (d),
       };
    \end{feynman}
  \end{tikzpicture}
\caption{}
\label{fig:feynmandiagram_pv_a}
\end{subfigure}
\begin{subfigure}[b]{0.32\textwidth}\centering 
\captionsetup{skip=0pt}
 \begin{tikzpicture}[baseline=(i.base)]
    \begin{feynman}[small]
      \vertex (i) ;
      \vertex [left = 1cm of i] (a) {\(D^*\)};
      \vertex [right = 1cm of i] (c) {\(D^*\)};
      \vertex [below = 2cm of i] (j);
      \vertex [below = 2cm of a] (b) {\(\bar{D}^*\)};
      \vertex [below = 2cm of c] (d) {\(\bar{D}^*\)};
      \diagram* {
        (a) -- [fermion] (i) -- [fermion] (c), 
        (i) -- [edge label={\(\rho,\omega,J/\psi\)}] (j),
        (b) -- [fermion] (j) -- [fermion] (d),
       };
    \end{feynman}
  \end{tikzpicture}
\caption{}
\label{fig:feynmandiagram_vv_a}
\end{subfigure}
\begin{subfigure}[b]{0.32\textwidth}\centering 
\captionsetup{skip=0pt}
   \begin{tikzpicture}[baseline=(i.base)]
    \begin{feynman}[small]
      \vertex (i) ;
      \vertex [above left = 1cm of i] (a) {\(D^*\)};
      \vertex [above right = 1cm of i] (c) {\(D^*\)};
      \vertex [below left = 1cm of i] (b) {\(\bar{D}^*\)};
      \vertex [below right = 1cm of i] (d) {\(\bar{D}^*\)};
      \diagram* {
        (a) -- [fermion] (i) -- [fermion] (c), 
        (b) -- [fermion] (i) -- [fermion] (d),
       };
    \end{feynman}
  \end{tikzpicture}
\caption{}
\label{fig:feynmandiagram_vv_b}
\end{subfigure}
 \caption{Diagrams of the interaction between pseudoscalar and vector mesons: (a) $t$-channel vector-meson exchange for ${\rm PV}\rightarrow {\rm PV}$, (b) $t$-channel vector-meson exchange for ${\rm VV}\rightarrow {\rm VV}$, and (c) four-vector contact term. Dashed lines depict pseudoscalar mesons and solid lines represent vector mesons.}
\end{figure}
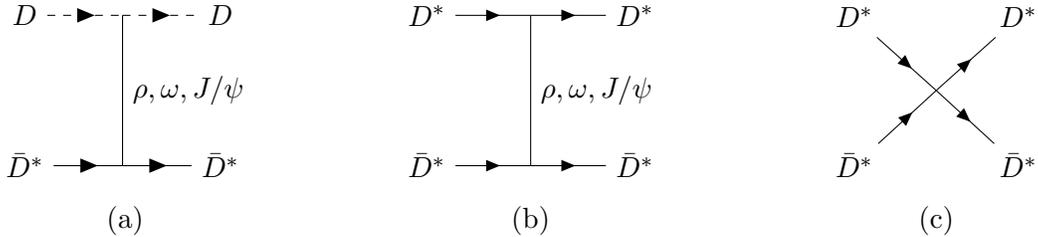

From the Lagrangians in Eqs.~(\ref{eq:LagVVVV})--(\ref{eq:LagVVV}) we build the $s$-wave amplitudes for the scattering processes $\rm{PV} \rightarrow {\rm PV}$ and ${\rm VV}\rightarrow {\rm VV}$. It turns out that in the hidden-gauge formalism the interaction between a pseudoscalar meson and a vector meson is described by a $t$-channel vector-meson exchange diagram, depicted in Fig.~\ref{fig:feynmandiagram_pv_a} for the case $D\bar{D}^*\rightarrow D\bar{D}^*$. For the interaction between two vector mesons, such as $D^*\bar{D}^*\rightarrow D^*\bar{D}^*$, in addition to the $t$-channel diagram of Fig.~\ref{fig:feynmandiagram_vv_a}, there is the contribution of the four-vector contact term of Fig.~\ref{fig:feynmandiagram_vv_b}. 

The VVV vertex of Eq.~(\ref{eq:LagVVV}) also allows one to build the $s$-channel and $u$-channel vector-meson exchange diagrams~\cite{Molina:2009ct,Molina:2010tx}. However, the first one only contributes to the $p$-wave interaction and thus it is suppressed, while the latter would imply the exchange of a virtual vector meson with double charm for the sector under consideration. Therefore these diagrams are left out in our study.

The interaction between bottomed mesons is described by the same Lagrangians as in the case of charmed mesons, with the replacement of the mesons with charm quarks in the $SU(4)$ matrices in Eqs.~(\ref{eq:matrixP}) and (\ref{eq:matrixVmu}) with the corresponding mesons with bottom quarks.
Therefore, in the hidden-bottom sector we have $B\bar{B}^*$ and $B_s\bar{B}_s^*$ coupled channels with $I^G(J^{PC})=0^+(1^{++})$, and $B^*\bar{B}^*$ and $B_s^*\bar{B}_s^*$ with $I^G(J^{PC})=0^+(2^{++})$.

\subsection{Pseudoscalar--=vector interaction}

We consider the $\{D\bar{D}^*,\,D_s\bar{D}_s^*\}$ ($\{B\bar{B}^*,\,B_s\bar{B}_s^*\}$) coupled-channel interaction, which proceeds through the exchange of virtual $\rho$, $\omega$, or $J/\psi$ ($\Upsilon$) mesons for diagonal transitions, as in Fig.~\ref{fig:feynmandiagram_pv_a}, or a virtual $K^*$ for nondiagonal transitions. The PPV and VVV vertices in this diagram are described by Eqs.~(\ref{eq:LagPPV}) and (\ref{eq:LagVVV}), respectively, and neglecting the three-momenta of the external vector mesons, which is a good approximation near threshold~\cite{Molina:2009eb}, they read
\begin{align}\label{eq:vertexPPV}
    t_{\textrm{PPV}}&= a\, (k_1+k_3)_i \epsilon_i^{(0)} \ , \\ \label{eq:vertexVVV}
    t_{\textrm{VVV}}&= b\, (k_2+k_4)_i \epsilon^{(2)}_j\epsilon^{(4)}_j\epsilon_i^{(0)} ,
\end{align}
where the indices $i,j$ are spatial and $\epsilon_i$ denotes the polarization three-vector components of the vector mesons. The superindex $(0)$ refers to the exchanged virtual vector meson and the indices $1,2,3,4$ denote the particles in the scattering process ${\rm P}(1)+{\rm V}(2)\rightarrow {\rm P}(3)+{\rm V}(4)$. The scattering amplitude is then obtained by considering all the possible vector-meson exchanges. The numerical factors $a$ and $b$ depend on the mesons in the vertex and are calculated in the charge basis.

In the limit $t\ll m_V^2$, with $t$ being the four-momentum exchanged in the process, the $t$-channel exchange diagram reduces to a contact interaction and the interaction potential from an incoming channel $l$ to an outgoing channel $m$ of the coupled-channel basis reads
\begin{equation}\label{eq:kernelPV}
 {V}_{lm}^{{\rm PV}\rightarrow {\rm PV}}(s,t,u)=\xi_{lm}(k_1+k_3)\cdot(k_2+k_4)\vec{\epsilon}\,^{(2)}\cdot\vec{\epsilon}\,^{(4)}=\xi_{lm}(s-u)\vec{\epsilon}\,^{(2)}\cdot\vec{\epsilon}\,^{(4)} \ ,
\end{equation}
where $s$, $t$, and $u$ are the usual Mandelstam variables. This amplitude is then transformed from the charge basis to the isospin basis and projected into spin states and into $s$ wave. We note that the approximation of neglecting the three-momenta of the external vector mesons with respect to their mass is appropriate only when considering the interaction projected in $s$ wave. Then, the only possible spin state with $L=0$ is $J=1$, for which the scalar factor $\vec{\epsilon}\cdot\vec{\epsilon}\,'=1$, up to corrections suppressed by the mass of the heavy-flavor meson. We have checked that, to the order we work here, the expression of the projected amplitudes obtained from the helicity formalism coincide with ours.

\begin{table}[htbp!]
\centering
\begin{tabular}{l | c}
\hline
 ${\rm PV}\rightarrow {\rm PV}$  & ${\xi}_{lm}^{(I=0)}$ \\
\hline
{$D\bar{D}^*\rightarrow D\bar{D}^*$}    & $-2g_D^2\left(\frac{1}{2m_{J/\psi}^2}+\frac{3}{4m_\rho^2}+\frac{1}{4m_\omega^2}\right)$ \\
{$D_s\bar{D}_s^*\rightarrow D_s\bar{D}_s^*$}   & $-2g_{D_s}^2\left(\frac{1}{2m_\phi^2}+\frac{1}{2m_{J/\psi}^2}\right)$    \\ 
{$D\bar{D}^*\rightarrow D_s\bar{D}_s^*$}   & $-2g_Dg_{D_s}\left(\frac{1}{\sqrt{2}m_{K^*}^2}\right)$ \\ 
\hline
\end{tabular}
\caption{The isospin coefficients of the $t$-channel vector-meson exchange terms for the channels involved in the generation of the $X(3872)$, with $I^G(J^{PC})=0^+(1^{++})$.}
\label{tab:xi_coeffPV}
\end{table}

The coefficients $\xi_{lm}$ in isospin basis are given in Table~\ref{tab:xi_coeffPV} in the hidden-charm sector for $I=0$. Those in the hidden-bottom sector are readily obtained with the replacement of the charmed mesons with the corresponding bottomed ones, and $J/\psi\rightarrow\Upsilon$. Also the constants $g_D$ and $g_{D_s}$ are replaced with $g_B$ and $g_{B_s}$.
One can see that the exchange of a heavy vector meson, $H=\{J/\psi,\,\Upsilon\}$, is suppressed with respect to the exchange of a light vector meson, $L=\{K^*, \rho,\omega,\phi\}$, by a factor $m_L^2/m_{H}^2$.

The interaction potential given in Eq.~(\ref{eq:kernelPV}) is equivalent to that obtained from the Lagrangian used by the authors of Ref.~\cite{Gamermann:2007fi},
\begin{equation}
 \mathcal{L}_{\textrm{PPVV}}=-\frac{1}{4f^2}{\textrm{Tr}}(J_\mu\mathcal{J}^\mu) \ ,
\end{equation}
with the pseudoscalar and vector currents defined as $J_\mu=(\partial_\mu P)P-P(\partial_\mu P)$ and $\mathcal{J}_\mu=(\partial_\mu V_\nu)V^\nu -V_\nu(\partial_\mu V^\nu)$, respectively, although employing another $SU(4)$-symmetry breaking pattern to suppress the terms involving the exchange of a heavy vector meson, which leads to somewhat different values for the diagonal ${\xi}_{lm}^{(I=0)}$ coefficients.
This interaction was recently used in Ref.~\cite{Cleven:2019cre} to study the thermal effects on the $X(3872)$.
In these works, the pion decay constant $f$ that holds in the light sector is replaced by that of the heavy $D$ meson, $f_D$, for channels involving heavy mesons.

Following a similar strategy, we take the hidden-gauge coupling constant $g$ in the Lagrangian of Eq.~(\ref{eq:LagHG}) to be channel dependent, i.e. $g_i\equiv \frac{m_\rho}{2f_i}$. The values of $f_i$ that we take for the charmed and bottomed mesons,
\begin{align}\label{eq:decayconstD}
 \sqrt{2}\,f_D&=212.6 \,{\rm MeV} \ ,\qquad \sqrt{2}\,f_{D_s}=249.9\,{\rm MeV} \ , \\
 \label{eq:decayconstB}
 \sqrt{2}\,f_B&=190.0 \,{\rm MeV} \ ,\qquad \sqrt{2}\,f_{B_s}=230.0\,{\rm MeV} \ ,
\end{align}
correspond to the preferred theoretical values of the most recent PDG review~\cite{pdg}.

One should bear in mind that small modifications of the amplitude in Eq.~(\ref{eq:kernelPV}) due to a different prescription for the hidden-gauge coupling constant, $g$, or for the $SU(4)$-breaking pattern, can be reabsorbed by the free parameters of the unitarization process, as it will become apparent in Sec.~\ref{sec:unitarization}. For instance, the value of the cutoff can be fixed to reproduce some experimental data. For the pseudocalar-vector interaction, we will fix the cutoff to reproduce the vacuum mass of the $X(3872)$.

\subsection{Vector-vector interaction}
\label{subsec:vv}
The vector-vector interaction in the hidden charm sector within the hidden-gauge formalism was addressed in~\cite{Molina:2009ct}. Here we consider the $\{D^*\bar{D}^*,\,D_s^*\bar{D}_s^*\}$ ($\{B^*\bar{B}^*,\,B_s\bar{B}_s^*\}$) coupled channels. Contrary to the pseudoscalar-vector case, the amplitude for the vector-vector scattering receives the contribution of a four-vector-meson contact term in Fig.~\ref{fig:feynmandiagram_vv_b} that can be directly obtained from the evaluation of the Lagrangian of Eq.~(\ref{eq:LagVVVV}):
\begin{equation}\label{eq:kernelVVc}
    {V}_{lm}^{{\rm VV}\rightarrow {\rm VV},\,\textrm{(c)}}=
    \alpha_{lm}\, \epsilon^{(1)}_i\epsilon^{(2)}_i\epsilon^{(3)}_j\epsilon^{(4)}_j
    +\beta_{lm}\,\epsilon^{(1)}_i\epsilon^{(2)}_j\epsilon^{(3)}_i\epsilon^{(4)}_j 
    +\gamma_{lm}\,\epsilon^{(1)}_i\epsilon^{(2)}_j\epsilon^{(3)}_j\epsilon^{(4)}_i \ .
\end{equation}

The contribution to the potential from the $t$-channel exchange is obtained in a similar way as in the pseudoscalar-vector case, except for the more complex structure of polarization vectors that follows from the evaluation of the diagram in Fig.~\ref{fig:feynmandiagram_vv_a} with the three-vector-meson vertex of Eq.~(\ref{eq:vertexVVV}),
\begin{equation}\label{eq:kernelVVex}
 {V}_{lm}^{{\rm VV}\rightarrow {\rm VV},\,\textrm{(ex)}}(s,t,u)=\tilde{\xi}_{lm}(s-u)\epsilon^{(1)}_i\epsilon^{(2)}_j\epsilon^{(3)}_i\epsilon^{(4)}_j \ .
\end{equation}

The amplitudes in Eqs.~(\ref{eq:kernelVVc}) and (\ref{eq:kernelVVex}) can be separated into three different spin contributions, $J=0,1,2$ with $L=0$ (we only consider the $s$ wave), by using the spin projectors $\mathcal{P}^{(J=0)},\,\mathcal{P}^{(J=1)},\,\mathcal{P}^{(J=2)}$~\cite{Molina:2008jw}. Given the following expressions for the combinations of polarization vectors,
\begin{align}
    \epsilon_i^{(1)}\epsilon_i^{(2)}\epsilon_j^{(3)}\epsilon_j^{(4)}&=3\mathcal{P}^{(J=0)} \ , \\
    \epsilon_i^{(1)}\epsilon_j^{(2)}\epsilon_i^{(3)}\epsilon_j^{(4)}&=\mathcal{P}^{(J=0)}+\mathcal{P}^{(J=1)}+\mathcal{P}^{(J=2)} \ , \\
    \epsilon_i^{(1)}\epsilon_j^{(2)}\epsilon_j^{(3)}\epsilon_i^{(4)}&=\mathcal{P}^{(J=0)}-\mathcal{P}^{(J=1)}+\mathcal{P}^{(J=2)} \ , 
\end{align}
one can see that the contact term
\begin{equation}\label{eq:kernelVVc-proj}
    {V}_{lm}^{{\rm VV}\rightarrow {\rm VV},\,\textrm{(c)}}= (3\alpha_{lm}+\beta_{lm}+\gamma_{lm})\mathcal{P}^{(J=0)}+(\beta_{lm}-\gamma_{lm})\mathcal{P}^{(J=1)}+(\beta_{lm}+\gamma_{lm})\mathcal{P}^{(J=2)}\ ,
\end{equation}
gives rise to interactions in all the spin channels with different weights, while the vector-meson exchange term,
\begin{equation}\label{eq:kernelVVex-proj}
 {V}_{lm}^{{\rm VV}\rightarrow {\rm VV},\,\textrm{(ex)}}(s,t,u)=\tilde{\xi}_{lm}(s-u)\left[\mathcal{P}^{(J=0)}+\mathcal{P}^{(J=1)}+\mathcal{P}^{(J=2)}\right] \ ,
\end{equation}
contributes in the same amount to all the spin states. 

With the above considerations, one has that the full expression for the ${\rm VV}\rightarrow {\rm VV}$ scattering amplitude in a sector with definite isospin $I$ and spin $J$ can be written as
\begin{equation}\label{eq:kernelVV}
 V_{lm}^{{\rm VV}\rightarrow {\rm VV},\,(I,J)}(s,t,u)=C_{lm}^{(I,J)}+\tilde{\xi}_{lm}^{(I,J)}(s-u) \ ,
\end{equation}
where the first term comes from the contact interaction and the second term corresponds to the $t$-channel vector-meson exchange. The corresponding isospin coefficients $C_{lm}^{(I,J)}$ and $\tilde{\xi}_{lm}^{(I,J)}$ are displayed in Table~\ref{tab:xi_coeffVV}, for $I=0$ and $J=2$, which are the quantum numbers that we assume for the $X(4014)$. The coefficients in the hidden-bottom sector follow from the same replacements as in the pseudoscalar-vector case. 
We have taken $g_{D^*/B^*}=g_{D/B}$ and $g_{D_s^*/B_s^*}=g_{D_s/B_s}$, with the values of $f_{D/B}$ and $f_{D_s/B_s}$ given above. 

\begin{table}[htbp!]
\centering
\begin{tabular}{l | c c}
\hline
 ${\rm VV}\rightarrow {\rm VV}$ & $C_{lm}^{(I=0,J=2)}$ & $\tilde{\xi}_{lm}^{(I=0,J=2)}$ \\
\hline
{$D^*\bar{D}^*\rightarrow D^*\bar{D}^*$}    & $-3g_D^2$ & $-2g_D^2\left(\frac{1}{2m_{J/\psi}^2}+\frac{3}{4m_\rho^2}+\frac{1}{4m_\omega^2}\right)$ \\
{$D^*_s\bar{D}_s^*\rightarrow D_s^*\bar{D}_s^*$}   &  $-2g_{D_s}^2$ & $-2g_{D_s}^2\left(\frac{1}{2m_\phi^2}+\frac{1}{2m_{J/\psi}^2}\right)$    \\ 
{$D^*\bar{D}^*\rightarrow D_s^*\bar{D}_s^*$}   &  $-\sqrt{2}g_Dg_{D_s}$ & $-2g_Dg_{D_s}\left(\frac{1}{\sqrt{2}m_{K^*}^2}\right)$ \\         
\hline
\end{tabular}
\caption{The isospin coefficients of the contact ($C_{ij}^{(I,J)}$) and the $t$-channel vector-meson exchange ($\tilde{\xi}_{ij}^{(I,J)}$) terms for the channels involved in the generation of the $X(4014)$, with isospin $I^G(J^{PC})=0^+(2^{++})$.}
\label{tab:xi_coeffVV}
\end{table}

It was shown in Ref.~\cite{Molina:2009ct} that the contribution of the contact term is in general smaller than the contribution of the vector meson exchange, but not negligible. For instance, in the $D^*\bar{D}^*\rightarrow D^*\bar{D}^*$ transition the contact term accounts for $\sim 5\%$ of the total interaction evaluated at the $D^*\bar{D}^*$ threshold for $I=0$, $J=2$ (see Table~XVII in~\cite{Molina:2009ct}).

It is also interesting to note that the contribution to the vector-vector amplitude from the dominant vector-meson exchange term in Eq.~(\ref{eq:kernelVV}), with any spin $J=0$, $1$, or $2$, has the same expression as the pseudoscalar-vector amplitude in Eq.~(\ref{eq:kernelPV}), with definite spin $J=1$, as is evident from the fact that $\xi_{lm}^{(I)}=\tilde{\xi}_{lm}^{(I,J)}$ in Tables~\ref{tab:xi_coeffPV} and \ref{tab:xi_coeffVV}. Hence, the leading interaction between open heavy-flavor mesons within the local-hidden gauge approach automatically fulfills the rules of HQSS. The contact term, only present in the vector-vector interaction, does not satisfy HQSS constraints but it is a numerically suppressed contribution~\cite{Fernandez-Soler:2015zhj}.



\section{Unitarization in coupled channels}
\label{sec:unitarization}
In this section we briefly discuss the unitarization process that leads to poles in the scattering amplitudes and that are associated to molecular states that are dynamically generated.
The pseudoscalar-vector and vector-vector interactions derived from the local hidden-gauge formalism are unitarized by solving the coupled-channel Bethe-Salpeter equation in the on-shell factorization approach~\cite{Oller:1997ti,Oset:1997it},
\begin{equation}\label{eq:BS}
 T=V(1-VG)^{-1} \ ,
\end{equation}
where the kernel $V$ is provided by the $s$-wave projection of the interaction potentials in Eqs.~(\ref{eq:kernelPV}) and (\ref{eq:kernelVV}), and $G$ is the propagator of the two mesons in the loop,
\begin{align}\label{eq:loop}\nonumber
    G(s)&=i\int \frac{d^4q}{(2\pi)^4}\frac{1}{q^2-m_1^2+i\varepsilon}\frac{1}{(P-q)^2-m_2^2+i\varepsilon} \\
    &=\int\frac{d^3q}{(2\pi)^3}\frac{\omega_1+\omega_2}{2\omega_1\omega_2}\frac{1}{s-(\omega_1+\omega_2)^2+i\varepsilon} \ ,
\end{align}
where $m_1$ and $m_2$ are the masses of the two open heavy-flavor mesons, $q$ is the four-momenta of one of the mesons in the loop, $P$ is the total four-momentum of the system, $P^\mu=(\sqrt{s},\vec{0})$, and $\omega_1=\sqrt{\vec{q}\,^2+m_1^2}$ and $\omega_2=\sqrt{\vec{q}\,^2+m_2^2}$ are the energies of the intermediate mesons. The integral in Eq.~(\ref{eq:loop}) needs to be regularized. There are two methods that are usually employed. One is the dimensional regularization scheme, in which the loop function reads
\begin{align}\label{eq:loopDR} \nonumber
 G^{\textrm{DR}}(s)=&\frac{1}{16\pi^2}\Big\{ a(\mu)+\ln\frac{m_1^2}{\mu^2}+\frac{m_2^2-m_1^2+s}{2s}\ln\frac{m_2^2}{m_1^2} \\  \nonumber
  &+\frac{q_{\textrm{cm}}}{\sqrt{s}}\left[ \ln\left(s-(m_1^2-m_2^2)+2q_{\textrm{cm}}\sqrt{s}\right)+\ln\left(s+(m_1^2-m_2^2)+2q_{\textrm{cm}}\sqrt{s}\right) \right.\\
  &\left. -\ln\left(-s+(m_1^2-m_2^2)+2q_{\textrm{cm}}\sqrt{s}\right)-\ln\left(-s-(m_1^2-m_2^2)+2q_{\textrm{cm}} \sqrt{s}\right) \right] \Big\} \ ,
\end{align}
where $q_{\textrm{cm}} (s)=\sqrt{[s-(m_1+m_2)^2][s-(m_1-m_2)^2]}/(2\sqrt{s})$ is the three-momentum of the mesons in the loop in the center-of-mass frame, and the subtraction constant $a(\mu)$ at the regularization scale $\mu$ is a free parameter. 
The other method consists in introducing a cutoff $\Lambda$ in the momentum integral shown explicitly after the last equality in Eq.~(\ref{eq:loop}).
In this case, $\Lambda$ is the free parameter. An analytical expression can be found in Ref.~\cite{Oller:1998hw}. 

The two regularization methods can be related by demanding the same value of the loop function at threshold, which results in a relationship between the free parameters $a(\mu)$ and~$\Lambda$:
\begin{equation}
    a(\mu)=16\pi^2\left[G^{\Lambda}(s_{\textrm{thr}})-G^{\textrm{DR}}(s_{\textrm{thr}})\right] \ ,
\end{equation}
for a given $\mu$.
A naturally sized cutoff in the range $[500,1000]\,\textrm{MeV}$ corresponds to values of the subtraction constant for the $D\bar{D}^*/D^*\bar{D}^*$ channels in the range $[-1.8,-2.4]$, for $\mu=1\,\textrm{GeV}$. We note that the cutoff method produces distortions in the loops at energies of a few hundreds of MeV above threshold, but it always gives $\textrm{Re\,}G<0$ below threshold, which is the region where bound states appear. We recall that bound states correspond to singularities in the scattering amplitude, which, according to Eq.~(\ref{eq:BS}), fulfill $1=V\,\textrm{Re\,}G$. Conversely, dimensional regularization can produce $\textrm{Re\,}G>0$ and the subsequent generation of unphysical states with repulsive potentials.

Following the previous observation, we have thus decided to use the cutoff regularization. On one hand, we are interested in bound states, where the use of a cutoff scheme is safer (to avoid generation of spurious poles). On the other hand, the calculation is straightforward to generalize to finite temperature. We will exploit the extension at finite temperature of the unitarization process with thermal propagators regularized with the cutoff that we presented in Refs.~\cite{Montana:2020lfi,Montana:2020vjg}, a methodology that is discussed in the next section.
The value of the cutoff is determined by fixing the pole position in vacuum to the experimental mass of the $X(3872)$ in the pseudoscalar-vector sector, and that of the $X(4014)$ in the vector-vector sector.

The unitarized scattering amplitude admits an expansion in a Laurent series around the pole position. For a $T$-matrix element it reads
\begin{equation}
    T_{ij}(z)=\frac{g_ig_j}{z-z_p}+\sum_{n=0}^{\infty}T_{ij}^{(n)}(z-z_p)^n \ ,
\end{equation}
where $z_p$ is the pole position in the complex $s$ plane, $g_i$ is the coupling of the bound state or resonance to channel $i$ (e.g., $i=D\bar{D}^*,D_s\bar{D}_s^*$), and $g_ig_j$ is the residue around the pole. We can get the couplings to each of the channels using the limit formula
\begin{equation}
    g_ig_j=\lim_{z\rightarrow z_p}\left[(z-z_p)T_{ij}(z)\right] \ ,
\end{equation}
or, alternatively,
\begin{equation}\label{eq:free-th-coupder}
g_ig_j=\Big[\frac{\partial}{\partial z}\Big(\frac{1}{T_{ij}(z)}\Big)\Big|_{z_p}\Big]^{-1} \ .
\end{equation}

\section{Finite temperature formalism}
\label{sec:temperature}

In Refs.~\cite{Montana:2020lfi,Montana:2020vjg} we employed the imaginary-time formalism (ITF) to compute the medium corrections due to a gas of light mesons on the properties of the open-charm pseudoscalar and vector mesons.
The general expression for the thermal two-meson propagator loop reads 
\begin{align}\label{eq:hot-loop-compact}\nonumber
 G(E,\vec{P}\,;T)  =\int\frac{d^3q}{(2\pi)^3}\int_{-\infty}^\infty d\omega\int_{-\infty}^\infty d\omega'&\frac{S_1(\omega,\vec{q}\,;T)S_2(\omega',\vec{P}-\vec{q}\,;T)}{E-\omega-\omega'+i\varepsilon}\\
 \times&\left[1+f(\omega,T)+f(\omega',T)\right] \ .
\end{align}
As mentioned in the previous section, the real part of the momentum integral is regularized with a cutoff, its value being determined in vacuum. The Bose-Einstein distribution factors $f(\omega,T)=1/(e^{\omega/T}-1)$ appear from the summation over Matsubara frequencies within the ITF.

The spectral functions $S_i(\omega,\vec{q};T)$ take account of the dressing of the heavy-meson masses by the thermal medium. 
We employ the spectral functions for the $D/D_s$ and $D^*/D_s^*$ mesons presented in Refs.~\cite{Montana:2020lfi,Montana:2020vjg}, and we have performed similar calculations to obtain those of the open-bottom mesons.

We then solve the Bethe-Salpeter equation in Eq.~(\ref{eq:BS}) with the above thermal loop functions containing dressed mesons, and the interaction kernels of Sec.~\ref{sec:hiddengauge}. We note that, in the ITF formalism, thermal corrections enter in loop diagrams~\cite{lebellac,kapustagale}, and therefore the tree-level interactions at finite temperature remain the same as in vacuum.

The analytical continuation of the unitarized scattering amplitudes to the complex-energy plane at finite temperature, with imaginary Matsubara frequencies, is nontrivial. Therefore, we do not look for poles of the scattering amplitude in the complex-energy plane at finite temperature. Instead, we extract the thermal properties of the dynamically generated states from the structures observed in $\textrm{Im}\,T$, as discussed in the next section.

\section{Results}
\label{sec:results}

We will apply the formalism described in the previous sections to discuss the plausibility of dynamically generating the $X(3872)$ and the $X(4014)$ as $D\bar{D}^*$ and $D^*\bar{D}^*$ molecular states, as well as the partners predicted in the bottom sector. After fixing the model in vacuum in order to reproduce the mass of these states from the experiments, we will analyze the modification of their properties in a hot medium.

\subsection{Vacuum properties}

Solving the Bethe-Salpeter equation in Eq.~(\ref{eq:BS}) with the hidden-gauge interaction kernel of Eq.~(\ref{eq:kernelPV}) for the $D\bar{D}^*$ and $D_s\bar{D}_s^*$ coupled-channel system with $I=0$ and $J=1$, we find a pole below the $D\bar{D}^*$ threshold that corresponds to a bound state and thus has no decay width. It can be placed at $m_{X(3872)}$ by using a cutoff value of $\Lambda=567$~MeV in the regularization of the loop functions, as shown in the first row of Table~\ref{tab:poles}. The generation of this state is dominated by the $D\bar{D}^*\rightarrow D\bar{D}^*$ interaction, as we can see from the large coupling of the pole to this channel. Actually, the $D_s\bar{D}_s^*$ channel plays a minor role, and its omission barely affects the pole position.

\begin{table}[htbp!]
 \centering
 \begin{tabular}{c c r@{}l r@{}l r@{}l r@{}l }
  \hline
  $\Lambda$  & State &  \multicolumn{2}{c}{Nearest threshold} &  \multicolumn{2}{c}{$\sqrt{z_p}$ }  & \multicolumn{4}{c}{Couplings} \\
   (MeV)  &  &  \multicolumn{2}{c}{(MeV)} & \multicolumn{2}{c}{(MeV)} & \multicolumn{4}{c}{(GeV)}  \\
  \hline
  $567$ & $X(3872)$ & $m_{D}+m_{\bar{D}^*}$&$\,=3875.80$ & $3871.65$&$\,+\,i\,0.00$ & $|g_{D\bar{D}^*}|$&$\,=9.23$ & $|g_{D_s\bar{D}_s^*}|$&$\,=3.98$ \\
  $510$ & $X(4014)$ & $m_{D^*}+m_{\bar{D}^*}$&$\,=4017.11$ & $4014.31$&$\,+\,i\,0.00$ & $|g_{D^*\bar{D}^*}|$&$\,=8.56$ & $|g_{D_s^*\bar{D}_s^*}|$&$\,=3.69$ \\
  $567$ & $-$ & $m_{B}+m_{\bar{B}^*}$&$\,=10604.12$ & $10548.65$&$\,+\,i\,0.00$ & $|g_{B\bar{B}^*}|$&$\,=7.51$ & $|g_{B_s\bar{B}_s^*}|$&$\,=3.28$ \\
  $510$ & $-$ & $m_{B^*}+m_{\bar{B}^*}$&$\,=10649.30$ & $10611.21$&$\,+\,i\,0.00$ & $|g_{B^*\bar{B}^*}|$&$\,=6.47$ & $|g_{B_s^*\bar{B}_s^*}|$&$\,=2.83$ \\
  \hline
 \end{tabular}
\caption{Pole position and coupling constants in the sectors with hidden charm (upper half) and hidden bottom (lower half), strangeness $S=0$ and $I^G(J^{PC})=0^+(1^{++}),0^+(2^{++})$, together with the value of the cutoff employed $\Lambda$, the label of the associated experimental state and the threshold of the nearest channel.}
\label{tab:poles}
\end{table}

Similarly, the unitarization of the interaction of Eq.~(\ref{eq:kernelVV}) for the $D^*\bar{D}^*$ and $D_s^*\bar{D}_s^*$ coupled-channel system with $I=0$ and $J=2$ gives rise to a pole in the scattering amplitude below the $D^*\bar{D}^*$  threshold that is placed at $m_{X(4014)}$ by taking $\Lambda=510$~MeV. 
Note that in this sector, in addition to the attractive vector exchange interaction, which is responsible for the generation of a bound $D\bar{D}^*$ state in the PV case, there is also the attractive contribution from the contact term. It is therefore not surprising that a lower value of the cutoff is needed in the VV sector to bind the $X(4014)$ a few MeV below the $D^*\bar{D}^*$ threshold.
We note that these values of the cutoff lie in the lower side of what is considered to be naturally sized.
The pole position of the $X(4014)$ and its couplings to the different channels are displayed in the second row of Table~\ref{tab:poles}.

The poles associated to the counterparts of the $X(3872)$ and the $X(4014)$ in the bottom sector are dynamically generated from the $B^{(*)}\bar{B}^*$ interaction employing, for consistency, the same values of the cutoff. The properties predicted for these states are listed in the subsequent rows of Table~\ref{tab:poles}. We obtain a $J=1$ $B\bar{B}^*$ state with a binding energy of $55$~MeV from the PV interaction, and a $J=2$ $B^*\bar{B}^*$ state with $38$~MeV of binding energy. Therefore, according to the model employed, these excited bottomonia states would be considerably more bound than those in the charm sector. This is partly related to having a somewhat stronger interaction in the bottom sector due to the slightly smaller decay constants [compare Eqs.~(\ref{eq:decayconstD}) and (\ref{eq:decayconstB})], but also to the smaller kinetic energy inherent in the bottomonia states because of the heavier masses of their constituents.

The authors of Ref.~\cite{Duan:2022upr} have recently suggested the $X(4014)$ to be a $D^*\bar{D}^*$ molecule with quantum numbers $I^G(J^{PC})=0^+(0^{++})$ using a contact interaction within the hidden-gauge formalism. The authors argue to have absorbed the vector-meson exchange contribution in the contact term by varying $g_D$ for values around $g_D=m_{D^*}/(2f_D)$. This is an arguable procedure considering that the contact term is subleading with respect to the meson-exchange contribution (see our discussion in Sec.~\ref{subsec:vv}), hence the latter should be the one determining the size and sign of the interaction. The adopted prescription results in a spin-$2$ interaction that is not attractive enough to dynamically generate a bound state at an energy close to the experimental mass of the $X(4014)$. However, by taking the full vector-vector interaction, the $X(4014)$ can be naturally generated as a $D^*\bar{D}^*$ bound state with $I^G(J^{PC})=0^+(2^{++})$, as we have shown in this section. 
Another argument that disfavors the $J=0$ assignment to the $X(4014)$ made in  Ref.~\cite{Duan:2022upr} is the fact that the contact interaction is repulsive in this sector, i.e. compelling the authors to employ small values of the subtraction constants in the unitarization process to produce an unphysical bound state.


We conclude the study at $T=0$ by providing an estimation of the size of the $X(3872)$ and $X(4014)$ from their wave functions, following the formalism of Ref.~\cite{Gamermann:2009uq}, which has been derived specifically for these exotics, thus using the nonrelativistic approach. After computing the wave functions in coordinate space as described by~\cite{Gamermann:2009uq}, the obtained values of the rms radius, defined as
\begin{equation}
r_{ \rm{rms}} = \sqrt{\langle r^2 \rangle} = \left( \int d^3r r^2 |\Psi|^2 \right)^{1/2} \ ,
\end{equation}
are $r_{\rm{rms}} [X(3872)]=1.96 \textrm{ fm}$ and $r_{ \rm{rms}} [X(4014)]=2.30 \textrm{ fm}$, which are typical sizes for shallow bound states (e.g. the deuteron). Within our description of the $X(3872)$ as a molecular state, our estimation of the radius lies in the low side when compared to other estimates~\cite{Esposito:2016noz,Braaten:2020iqw}. The reason is that we work within the isospin formalism, and our vacuum meson masses correspond to an average of the charged states ($D^+,D^0,...$). Then, despite having fixed the exotic masses to their experimental positions, the resulting elastic thresholds and binding energies differ from those using charged states, cf. Table~\ref{tab:poles}. In fact, our result is a weighted average between the low binding energy ($\sim 0.1$ MeV) in the $D^0 \bar{D}^{*0}$ channel and the large binding ($\sim 8$ MeV) in the $D^+D^{*-}$ channel found in Ref.~\cite{Gamermann:2009uq}.

\subsection{Thermal modifications}

\begin{figure}[htbp!]
\includegraphics[width=\textwidth]{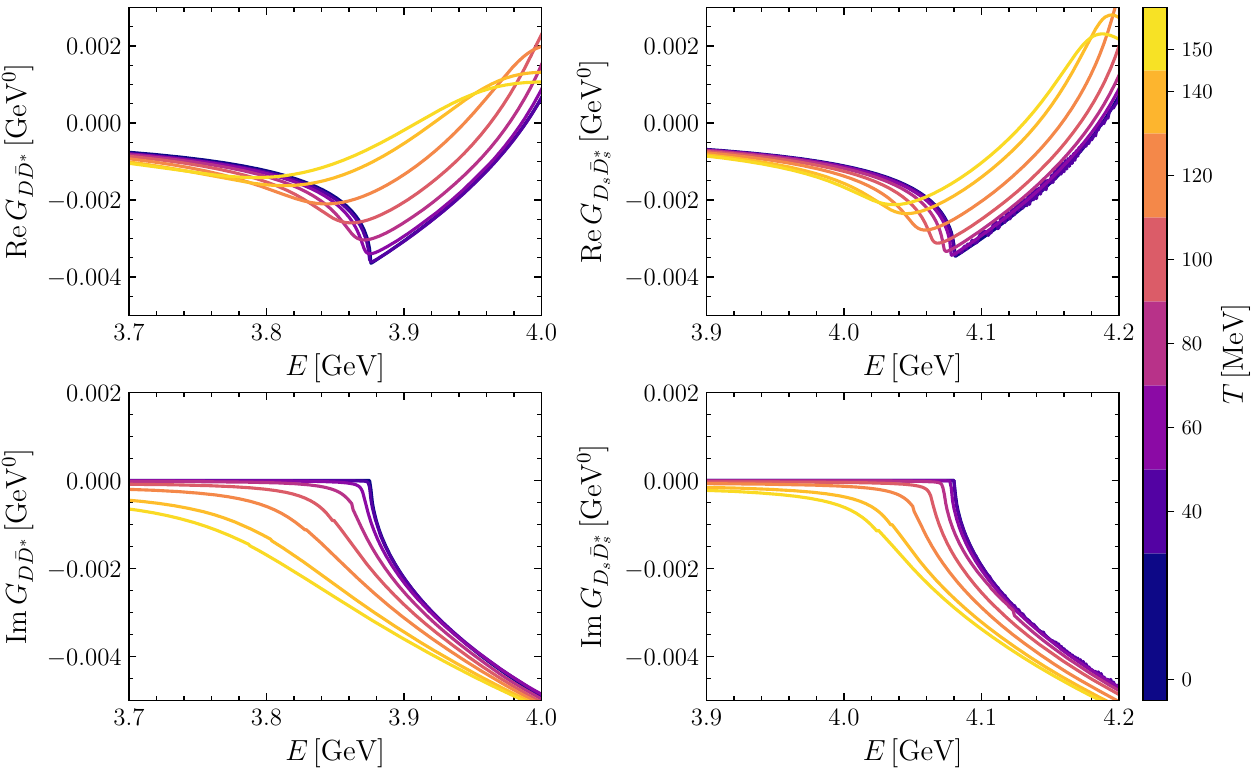}
\caption{Real (top panels) and imaginary (bottom panels) parts of the loop functions of the $D\bar{D}^*$ (left panels) and $D_s\bar{D}_s^*$ (right panels) channels at temperature $T$.}
\label{fig:loops_charm_pv}
\end{figure}

In this section we study how the dynamically generated $X(3872)$ and $X(4014)$ states, as well as their bottom counterparts, behave in a hot medium. 

We first show the thermal two-meson loop function of the $D\bar{D}^*$ (left panels) and $D_s\bar{D}_s^*$ channels (right panels) at finite temperature in Fig.~\ref{fig:loops_charm_pv}. These follow from the numerical integration of Eq.~(\ref{eq:hot-loop-compact}), where we have used the spectral functions of the $D$, $D_s$, $D^*$ and $D_s^*$ mesons obtained in our previous works~\cite{Montana:2020lfi,Montana:2020vjg}. In these references it was found that the thermal masses (i.e. the peak position of the spectral functions) decrease with increasing temperatures, while their widths grow considerably (see Fig.~\ref{fig:MassWidthGS} below). Therefore, the dressing of the loop function with the spectral functions softens and shifts towards lower energies the onset of the unitary cut of the imaginary part (bottom panels) with the consequent effects on the corresponding structure of the real part (top panels).



\begin{figure}[htbp!]
\begin{subfigure}[b]{0.47\textwidth}\centering 
\includegraphics[width=\textwidth]{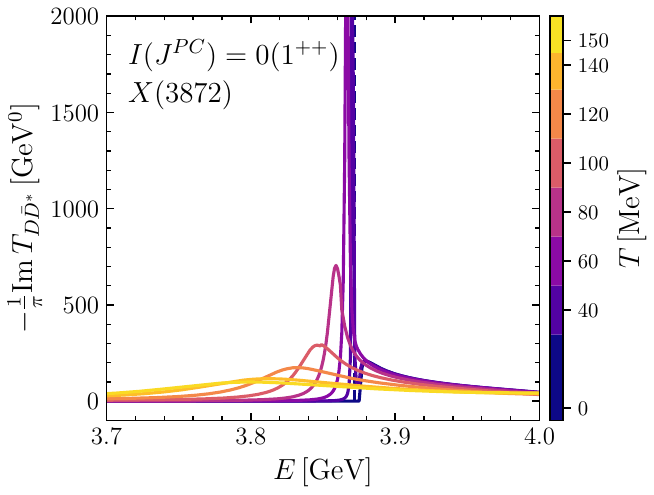}
\end{subfigure}
\begin{subfigure}[b]{0.47\textwidth}\centering 
\includegraphics[width=\textwidth]{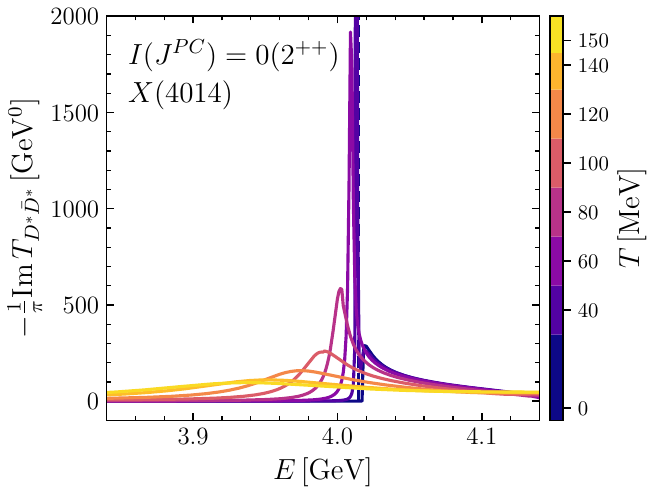}
\end{subfigure}
\caption{Thermal spectral function of the $X(3872)$ (left panel) and $X(4014)$ (right panel) states, i.e., imaginary part of the unitarized scattering amplitudes at finite temperature for the diagonal transitions $D \bar{D}^* \rightarrow D \bar{D}^*$ (left) and $D^* \bar{D}^* \rightarrow D^* \bar{D}^*$ (right).}
\label{fig:tmatrixX}
\end{figure}

\begin{figure}[htbp!]
\begin{subfigure}[b]{0.47\textwidth}\centering 
\includegraphics[width=\textwidth]{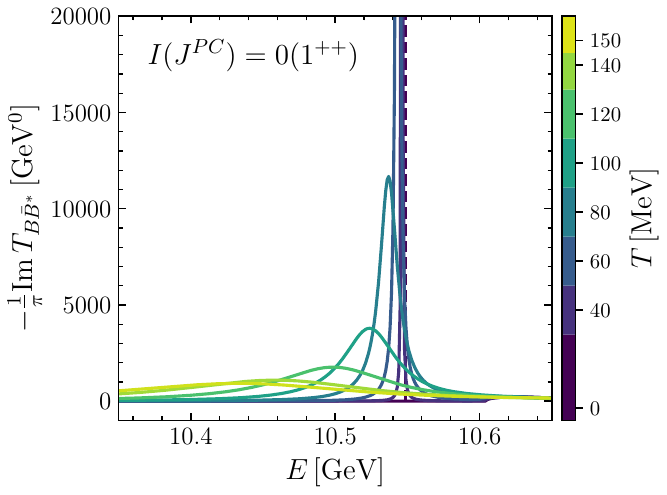}
\end{subfigure}
\begin{subfigure}[b]{0.47\textwidth}\centering 
\includegraphics[width=\textwidth]{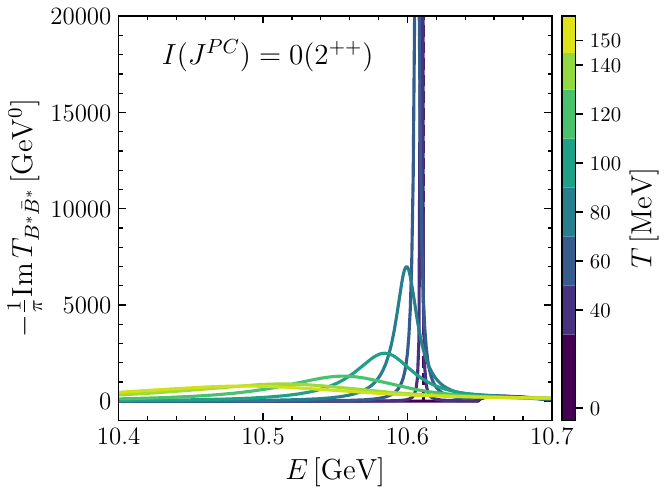}
\end{subfigure}
\caption{Thermal spectral function of the $X_b$ vector (left panel) and tensor (right panel) generated states, i.e., imaginary part of the unitarized scattering amplitudes at finite temperature for the diagonal transitions $B\bar{B}^* \rightarrow B\bar{B}^*$ (left) and $B^*\bar{B}^* \rightarrow B^*\bar{B}^*$ (right).}
\label{fig:tmatrixXb}
\end{figure}

In Fig.~\ref{fig:tmatrixX} we display the spectral functions of the dynamically generated states, $X(3872)$ (left panel) and $X(4014)$ (right panel), for various temperatures. Being dynamically generated states, these spectral functions are defined as
\begin{equation}
\label{eq:spec_def}
{\cal S}(\vec{P}=0, E)=-\frac{1}{\pi}\,{\rm Im\,} T(\vec{P}=0, E) \ ,
\end{equation}
where $T$ is the unitarized thermal scattering amplitude $D\bar{D}^*\rightarrow D\bar{D}^*$ (left panel) and $D^*\bar{D}^*\rightarrow D^*\bar{D}^*$ (right panel).  The dashed lines in dark blue depict the position of the poles associated to the $X(3872)$ and the $X(4014)$ in vacuum. As temperature increases, the peak position decreases and the width increases. 
The spectral functions of the dynamically generated $X_b$ states in the bottom sector, displayed in Fig.~\ref{fig:tmatrixXb}, show an analogous behavior with temperature as that of their charm counterparts.

\begin{figure}[h!]
\includegraphics[width=0.9\textwidth]{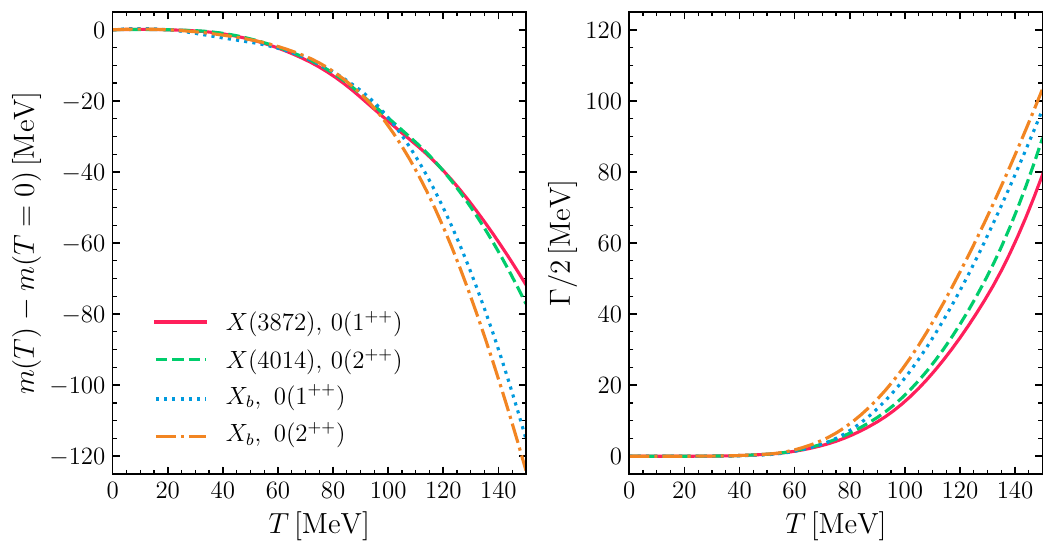}
\caption{Temperature evolution of the mass shift and the half-width of the hidden-charm and hidden-bottom dynamically generated states.}
\label{fig:MassWidth}
\end{figure}

Figure~\ref{fig:MassWidth} shows the evolution with temperature of the properties of the dynamically generated states. The mass shift represented on the left panel is obtained, at each temperature, as the difference between the peak position of the corresponding spectral function and the real part of the vacuum pole listed in Table~\ref{tab:poles}. The half width shown in the right panel is taken as half of the full width at half maximum from the left-hand side of the spectral distribution, as it is less distorted than the right-hand side which can be affected by threshold effects.
For comparison, we display in Fig.~\ref{fig:MassWidthGS} the thermal modification of the properties of the ground-state $D$, $D^*$, $\bar{B}$, and $\bar{B}^*$ mesons, extracted from their corresponding spectral functions.\footnote{The interaction of open heavy-flavor mesons with a bath of light mesons with vanishing net flavor chemical potentials is no different for heavy mesons or heavy anti-mesons, meaning that the in-medium spectral functions for $\bar{D}^{(*)}$/$B^{(*)}$ are the same as those for $D^{(*)}$/$\bar{B}^{(*)}$.}

\begin{figure}[h!]
\includegraphics[width=0.87\textwidth]{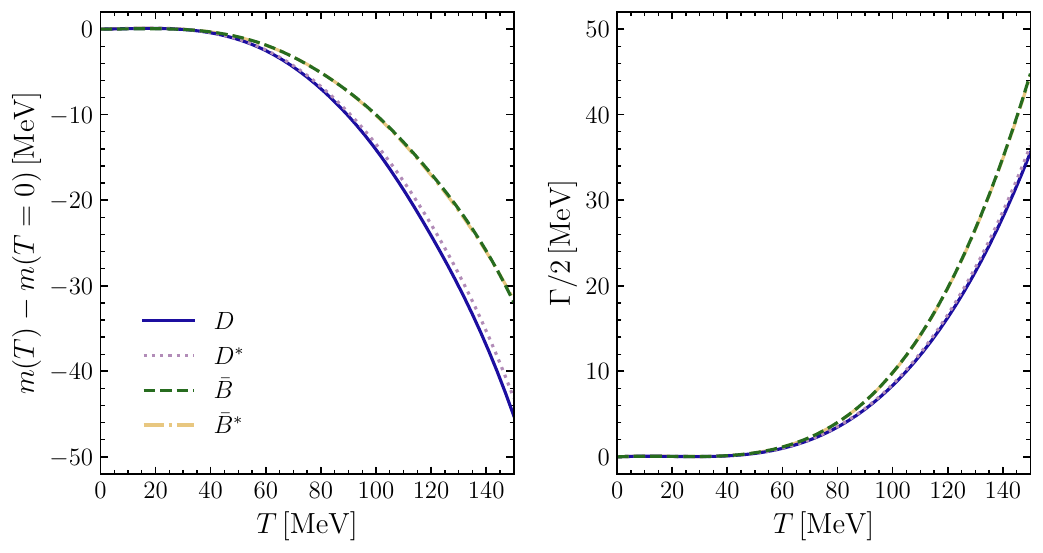}
\caption{Temperature evolution of the mass shift and the half-width of the open-charm and open-bottom ground state mesons.}
\label{fig:MassWidthGS}
\end{figure}

\begin{figure}[htbp!]
\includegraphics[width=0.9\textwidth]{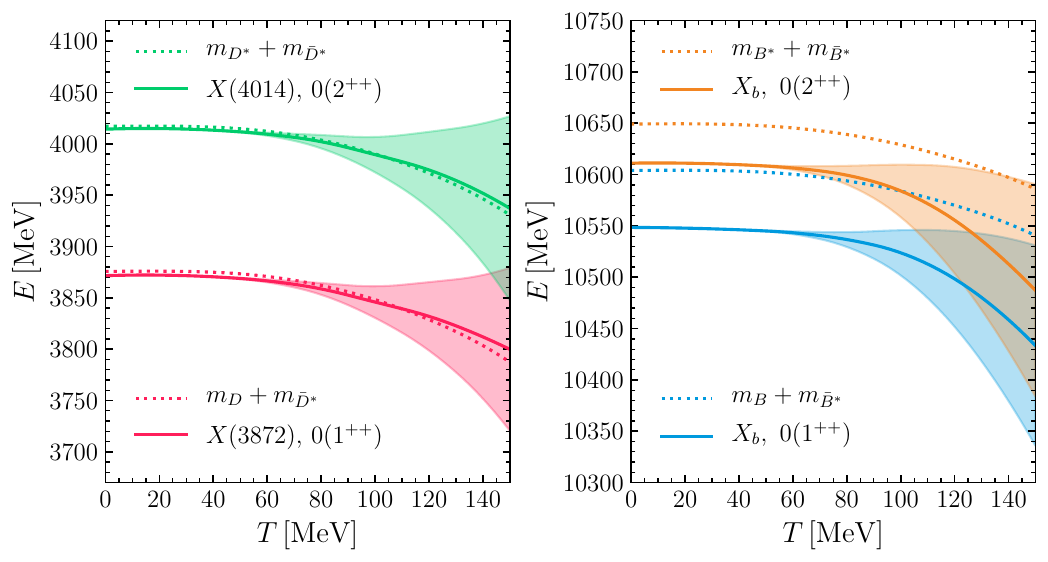}
\caption{Temperature evolution of the peak of the dynamically generated states (solid lines) and the thresholds (dotted lines). The shaded areas show the width of the states, e.g. $m_{X(3872)}\pm \Gamma_{X(3872)}/2$ for the $X(3872)$. }
\label{fig:thresholds}
\end{figure}

Focusing first on the hidden charm states, $X(3872)$ and $X(4014)$, we observe that their mass gets reduced with temperature, following the decreasing mass trend of their meson components. The mass reduction is around $30$~MeV at $T=100$~MeV and around $70$~MeV at $T=150$~MeV.  Nevertheless, the mass of the dynamically generated states is always close to their corresponding finite temperature $D \bar{D}^*$ and $D^*\bar{D}^*$ threshold,  respectively, being below it for $T \lesssim 100$~MeV and above it for $T\gtrsim 100$~MeV. This is more clearly visualized in the left panel of Fig.~\ref{fig:thresholds}, where the temperature evolution of the $X(3872)$ and $X(4014)$ masses (solid lines) is compared with their corresponding temperature dependent threshold (dashed lines).
The shaded areas account for the width of these states, which is found to be roughly the sum of the width of their meson components, being just somewhat larger for the highest temperatures explored in this work.
This can be easily checked upon comparing the half widths of the $D$ and $D^*$ mesons displayed on the right panel of Fig.~\ref{fig:MassWidthGS} with those of the  $X(3872)$ and $X(4014)$ states shown on the right panel of Fig.~\ref{fig:MassWidth}.

The displacement of the $X(3872)$ towards lower energies with increasing temperatures found in the present work is opposite to what is found in the study of Ref.~\cite{Cleven:2019cre}. There, the peaks of the $D$ and $D^*$ spectral functions were kept at the position of their vacuum mass, so a temperature-independent $D \bar{D}^*$ threshold was maintained. Therefore, the tendency of the dynamically generated state to move towards the threshold, and even above it,  with increasing temperature led the study of Ref.~\cite{Cleven:2019cre} to obtain an increased temperature-dependent $X(3872)$ mass with respect to its vacuum value. It is interesting to note that a recent lattice QCD study of temperature effects on charmed pseudoscalar and vector correlators finds a reduction in mass of 20(7)~MeV for the $D$ meson and of 43(10)~MeV for the $D^*$ meson~\cite{Aarts:2022krz}, which is quite consistent with our effective field theoretical results.

As for the temperature dependence of the $X(3872)$ width, while we find agreement with the value of about $30$~MeV at $T= 100$~MeV quoted in Ref.~\cite{Cleven:2019cre}, a large discrepancy of more than a factor of 2 is observed at $T=150$~MeV. At this temperature we find the $X(3872)$ to have a width of about $150$~MeV to be compared with the value of $60$~MeV given in  Ref.~\cite{Cleven:2019cre}.  This reduced width for the $X(3872)$ can partly be explained by the smaller width of its meson $D$ and $D^*$ components, found to be of the order of $45$~MeV at $T=150$~MeV in the model employed in Ref.~\cite{Cleven:2019cre}. We also note that the width of the $X(3872)$ depends on how it is extracted from an asymmetric peak which is distorted by a threshold. Considering the right-hand side of the distribution displayed in Ref.~\cite{Cleven:2019cre}, a width of $90-100$~MeV at $T=150$~MeV is obtained for the  $X(3872)$, a value that is more consistent with the width of its constituents and does not deviate so strongly from our result.

We end this section by discussing the properties of the hidden bottom states. 
As found in the charm sector, the mass of the dynamically generated $X_b(1^{++})$ and  $X_b(2^{++})$ states, displayed on the left panel of Fig.~\ref{fig:MassWidth}, decreases with temperature following the trend of the mass of their meson components seen on the left panel of Fig.~\ref{fig:MassWidthGS}. It is worth mentioning here that the temperature correction to the mass of the ground-state bottom mesons is found to be somewhat smaller than that of the charm mesons. Recalling that the mass shift is proportional to $\Pi/2M$, this behavior results from the combination of a twice more attractive self-energy~\cite{2022_Gloria} with a three times larger mass of the bottom mesons compared to the values for the charm mesons. 
The mass reduction of the $X_b(1^{++})$ and  $X_b(2^{++})$ bottomonia with respect to their vacuum mass is similar to that of their charmonia counterparts up to $T=100$~MeV and becomes larger at higher temperatures, reaching a value of around $120$~MeV at $T=150$~MeV, which is almost twice the reduction observed for the  $X(3872)$ and $X(4014)$ states. 
As clearly visualized in Fig.~\ref{fig:thresholds}, within the temperature range explored, the  bottomonia maintain a similar binding energy with respect to the temperature-dependent threshold as that in vacuum, although a stronger descent is observed at larger temperatures,
where one is also approaching the limit of applicability of hadronic effective theories of $T\sim T_c$. As for the width of the  $X_b(1^{++})$ and  $X_b(2^{++})$ states, displayed on the right panel of Fig.~\ref{fig:MassWidth}, it is found to be larger at each temperature than that of their charm counterparts, reaching a value of 50~MeV at $T=100$~MeV and about $200$~MeV at $T=150$~MeV.  This is obviously due to the wider character of their ground-state bottom meson components compared to the charm meson ones, as can be seen on the right panel of Fig.~\ref{fig:MassWidthGS}.

\section{Conclusions}
\label{sec:conclusions}

In this study we have addressed both vacuum and finite-temperature properties of the exotic state $X(3872)$ and its heavy-quark spin-flavor partners, namely, the $X(4014$) as well as the axial-vector and tensor $X_b$ states. For these goals we have assumed that the internal structure of these hadrons is of a meson-meson molecular state~\cite{Tornqvist:2004qy,Wong:2003xk,Thomas:2008ja,Gamermann:2009fv, Gamermann:2009uq,Wang:2013daa,Abreu:2016qci,ExHIC:2017smd}. While a tetraquark component has not been discarded, we based our assumption on the closeness of the mass of the $X(3872)$ to the $D \bar{D}^*$ threshold and its very small decay width. Using an effective hadron approach based on the hidden-gauge Lagrangian extended to $SU(4)$ and coupled to pseudoscalar mesons, we have first generated the $X(3872)$ out of the solution of the Bethe-Salpeter equation that resums the $s$-wave scattering of a charm and anticharm meson pair. After fixing the parameters of the model using the experimental properties of the $X(3872)$, we have extended the description to finite temperature using the imaginary-time formalism. Additional input for this calculation was the known thermal spectral functions of the open-charm mesons from Refs.~\cite{Montana:2020lfi,Montana:2020vjg}. Then, we have extended the calculation (both in vacuum and at finite temperature) to generate the tensor state, identified with the experimental $X(4014)$, out of the scattering of vector mesons. Finally we have studied their bottom counterparts $X_b$ in the axial-vector $I (J^{PC})=0 (1^{++})$ and tensor channels $I (J^{PC})=0 (2^{++})$, states which have not been experimentally found yet~\cite{CMS:2013ygz,ATLAS:2014mka,Belle:2014sys,Belle-II:2022xdi}, and for which we provide predictions, also at finite temperatures.


In our approach we have considered temperatures below the hadronization one, around $T_c=156$~MeV. A temperature of $T\simeq 150$~MeV roughly corresponds to our limiting temperature where our results start to suffer from sizable systematic uncertainties. For somewhat lower temperatures (corresponding to the freeze-out one) our results show that the mass of the $X(3872)$ and the $X(4014)$ would drop $\Delta m \simeq -60$~MeV, and acquire a decay width of $\Gamma \simeq 120$~MeV. While the small relative mass drop might not affect appreciably the dynamics of the exotics, the amount of decay width can give a somewhat larger multiplicity of these particles in the medium. A simple estimation of Boltzmann multiplicities gives an increased yield from the case using a vacuum mass (like the one used in the statistical thermal model) to a state with a downshifted mass and a decay width like the one obtained here. Such a moderate decay width is mostly inherited from the thermal width of the $D/D^*$ mesons. In principle this would point out to a higher multiplicity of open-charm states coming from the decays of the $X(3872)/X(4014)$. 

In vacuum, we find a similarity between these states and the deuteron case, as their binding energies and sizes are alike. Then, because their small binding energy,  the survival of these states in a hot environment could be questioned along the lines of the so-called ``snowballs in hell'' puzzle~\cite{Oliinychenko:2018ugs}. However notice that at $T \simeq 140$ MeV our exotics are genuine resonances (see Figs.~\ref{fig:MassWidth} and \ref{fig:thresholds}) with a decay width $\Gamma \simeq 120$ MeV, and a corresponding lifetime around $c\tau \simeq 1.6$ fm. Therefore, we would expect that these states suffer from subsequent dissociation and regeneration processes within the hadron phase, until their decoupling from the medium. A real-time simulation---like the ones performed in Ref.~\cite{Abreu:2016qci} but with broad states---is needed to ascertain the possibility of increased open charm production. Such continuous melting/fusion reactions have been considered for the deuteron case in several recent works~\cite{Oliinychenko:2018ugs,Staudenmaier:2021lrg, Neidig:2021bal}.

Another alternative to study these exotics might come from femtoscopic measurements in proton-proton and HICs~\cite{Lisa:2005dd,Fabbietti:2020bfg}. For the former, some works have seen the effect of a narrow (bound state) exotic in correlation functions~\cite{Kamiya:2022thy}. Should these measurements be feasible in HICs, it could be possible to study how the effect of having an in-medium resonance, instead of a bound state, can show up in the correlation function of $D$ and $D^*$ mesons.

Finally, within the bottom sector we found that---in line with the open-flavor meson studies in~\cite{Montana:2020lfi,Montana:2020vjg}---medium modifications affect heavier states more. While $X_b$ states have larger vacuum masses, the mass shifts at our top temperatures can be a factor of 2 larger than those for the $X$ states. In addition, we find slightly larger thermal decay widths for the bottom exotics. One should keep in mind that, as these states are much more suppressed in HICs than the charmed ones, the possible detection of any medium modification is really a far-reaching result. Nevertheless, there exist also theoretical works and lattice-QCD calculations exploring the melting of heavy mesons at finite temperature in connection to the chiral symmetry restoration. Although lattice-QCD calculations of hidden bottom states are typically focused on the bottomonium ground and excited states $\Upsilon$~\cite{Rothkopf:2019ipj}, similar studies addressing bottom exotics could in principle be also performed. In our case we predict that, under the assumption of the $X_b$'s being molecular states of $B\bar{B}^*/B^* \bar{B}^*$, they acquire thermal decay widths of $\Gamma \simeq 200$ MeV and mass shifts of $\Delta m \simeq - 120$ MeV for temperatures around $T = 150$ MeV.
\\

\begin{acknowledgments}
We thank V. Mathieu for checking the consistency of our partial wave amplitudes with those obtained within the helicity formalism.
This research has been supported from the Projects No. CEX2019-000918-M, No. CEX2020-001058-M (Unidades de Excelencia ``Mar\'{\i}a de Maeztu"), No. PID2019-110165GB-I00, and No. PID2020-118758GB-I00, financed by the Spanish Grant No. MCIN/AEI/10.13039/501100011033, as well as by the EU STRONG-2020 project, under the program H2020-INFRAIA-2018-1 Grant Agreement No. 824093. G.M. acknowledges support from the FPU17/04910 Doctoral Grant from the Spanish Ministerio de Universidades, and U.S. DOE Contract No. DE-AC05-06OR23177, under which Jefferson Science Associates, LLC, operates Jefferson Lab. L.T. and J.M.T.-R. acknowledge support from the DFG through Projects No. 411563442 (Hot Heavy Mesons) and No. 315477589-TRR 211 (Strong-interaction matter under extreme conditions). L.T. also acknowledges support from the Generalitat Valenciana under Contract No. PROMETEO/2020/023.

\end{acknowledgments}

\bibliographystyle{utphys}   
\bibliography{heavy-heavy}

\end{document}